\documentclass[aps,prb,preprint]{revtex4}
\usepackage{graphicx}
\usepackage{amsmath}
\usepackage{scalerel,stackengine}
\stackMath
\newcommand\reallywidehat[1]{%
\savestack{\tmpbox}{\stretchto{%
  \scaleto{%
    \scalerel*[\widthof{\ensuremath{#1}}]{\kern-.6pt\bigwedge\kern-.6pt}%
    {\rule[-\textheight/2]{1ex}{\textheight}}
  }{\textheight}%
}{0.5ex}}%
\stackon[1pt]{#1}{\tmpbox}%
}

\begin{document}

\title{Commentary on Peierls's ``In defence of ``measurement"" \cite{Peierls1991}.}

\author{Tuck C. Choy}
\email{tuckvk3cca@gmail.com}

\affiliation{Samy Maroun Center for Space, Time and the Quantum, Parc Maraveyre B${\hat a}$t. 1, 13260, Cassis, Bouches du Rh${\hat o}$ne, France  and \\
Departmento de Fisica-CIOyN, Universidad de Murcia, Murcia 30071, Spain.}


\begin{abstract}

Commentary prepared to accompany the reprint article by Rudolph Peierls\cite{Peierls1991} in the centenary volume for the celebration of 100 years of Heisenberg's quantum mechanics, to be published by World Scientific in July 2025.

\end{abstract}

\maketitle


\vskip 1 cm

\begin{quote}

``I do not know what's reality, that is a question for metaphysics, not physics."

{\rm Rudolph Peierls in private discussions with the author 1992.}

\end{quote}

\noindent
This year's centenary celebrations for the founding of quantum mechanics by Werner Heisenberg in 1925 also coincides with the thirtieth anniversary of the death of Rudolph Peierls in 1995. Peierls was Heisenberg's student, perhaps his most illustrious one, well known to be the father of the atomic bomb, see Frank Close \cite{FrankClose1}. In writing this commentary on Peierls's paper of Physics World in 1991 \cite{Peierls1991} I thought it might be useful to recall some memories and discussions I had with Rudy, as we called him in Oxford, pertaining in particular to fundamental questions in quantum mechanics, as a tribute to both men and as a prelude to my views on his response to Bell \cite{Bell1990}. Heisenberg was of course famous for his discovery of matrix mechanics in 1925 and as a student Peierls had learnt a lot from him and his way of doing physics and also from Pauli who had many anecdotes which Peierls recalled to me on occasions over the dinner table. Concerning quantum mechanics, I was always interested in discussions with him on the Landau and Peierls (1930) \cite{Peierls-Landau1930} paper on the uncertainty in the measurement of fundamental quantities imposed by relativity (finite speed of light $c$), which is the Landau-Peierls expression for the relativistic Heisenberg uncertainty principle: $\Delta p \Delta t \sim \hbar/c$. This would imply amongst other things that a precise measurement for the momentum say of the electromagnetic field i.e. the photon (which by default is a relativistic object) would require an infinite amount of time. This result came as such a shock to Niels Bohr that both young men, Peierls and Landau, were detained in Copenhagen in 1930, unable to leave until Bohr was satisfied that a resolution has been found. In fact, it was not until 1950 that Bohr and Rosenfeld finally published their paper \cite{Bohr-Rosenfeld 1950}, which was a masterpiece on measurement analysis, which Bohr insisted must be conducted with classical apparatus. The psychological effects on both Peierls and Landau must have been deep and also on Heisenberg who later became involved \cite{Bohr-Rosenfeld 1950}. Heisenberg later changed his own earlier views on quantum mechanics and adopted the S matrix theory for particle interaction processes \cite{Heisenberg1938}, as the only description that has observable physical meaning. As for Peierls, my feeling was that he was completely broken down by Bohr (a very persistent and persuasive man) while Landau may not have been so. Landau never accepted quantum field theory as a logically consistent theory due to infinities but partly due to the Landau-Peierls results of 1930 \cite{Peierls-Landau1930,Landau-Lifshitz4}. Prior to this they had also together penned their paper on quantum electrodynamics in configuration space, proposing for the first time a Schr\"odinger equation for light quanta \cite{Peierls-Landau21930}. The paper inspired Enrico Fermi to use their methodology for the theory of beta decay.

I only met John Bell once, in 1987 at the Schr\"odinger centenary at Imperial College London; his paper on the Ghiradi-Rimini-Weber (GRW) model,  which preceded CN Yang's did not leave a mark in my memory. However later conversations with Peierls taught me a few things about Bell, who was a student at Birmingham in 1956 while Peierls was the Head of Department.  He was a clever student, difficult at times as they all are, but had deep insight in all aspects of physics. I recall also that my ex-boss and mentor Peter Schofield (while I was a research scientist at the Atomic Energy Authority (AERE) at Harwell in 1988) telling me about Walter Marshall who was also a brilliant ex-student of Peierls's and a contemporary of Bell's. Walter later became director of the AERE in 1968 and then Chairman of the UK Atomic Energy Authority (UKAEA) in 1981.  Peter told me that Marshall came to him one day in tears exasperated saying: ``How come whenever I get into an argument with Bell, he is always right and I am wrong?". As to quantum mechanics Peierls, must have had a tough time teaching Bell anything concerning the status quo. Peierls told me that Bell had actually thought that experiments would show that he was right i.e. they would not violate his inequalities, but Aspect's experiments, and later, those of others had left him rather disappointed. Perhaps Bell's later devotion to the GRW model with its hidden variables and unconventional probability rules was his desperate effort to see the quantum world his way.  The above quote from Peierls came out of one of my discussions with him on these issues.  So now we can go on to the paper.

For a start, Bell's 1990 article actually disappointed me in later years. I have simply no idea why in his critique of ``Against Measurement" \cite{Bell1990} he chose three books: 1) Landau and Lifshitz's Quantum Mechanics,  2) Kurt Gottfried's Quantum Mechanics, and 3) Nicole van Kampen's Ten theorems about quantum mechanical measurements. I would have thought he should have gone for the jugular and tear to pieces (1) Dirac's : Principles of Quantum Mechanics, (2) Heisenberg's : Physical Principles of the Quantum theory  and (3) Herman Weyl's : The theory of Groups and Quantum Mechanics.  Those would be my three good books if I am daring enough to challenge the principles of quantum mechanics.  Dirac never got involved in wavefunction collapse or the shifty split.  The closest he got to that would be section 10 in his book Principles of Quantum Mechanics in which he assumed that if a measurement is made, “the disturbance causes a jump in the state of the dynamical system” .  I do not however buy Bell’s charge that Dirac was the most distinguished of the `why bother?'ers.  For others yes, but for Dirac, it was not the case.  Bell himself had mitigated his own charge by saying that Dirac acknowledged that his class I type problems exist but the time was not ripe to be able to go into them. Again in his 1963 Scientific American paper\cite{Dirac1963} for example he considers that a consistent picture for quantum mechanics (of the type Bell was looking for) is a too hard Class I type problem. He had in fact worried a lot about quantum mechanics foundations and proposed that $\hbar$ may not stand as firm as $e$ and $c$ in the fine structure constant $\alpha = {e^2/\hbar c}$ and could change.  He stubbornly refused to accept the Renormalization program  as the final word, but in this even Richard Feynman was the same. Modern particle physics vindicated Dirac's concerns to some extent as, at sufficiently high energies, it was found that the value of $\alpha$, now known to 8 or 9 decimal places does indeed change.

In Peierls's article two things stand out.  In such a short two page article Peierls emphasized the uncertainty principle four times and this is for him the very basis for the foundation and the formalism of quantum mechanics, not the other way round.  In this he was true to the spirit of Heisenberg. Secondly he adopts the van Kampen stand that the wavefunction or density matrix represents our knowledge of the system. It has no other physical meaning beyond that. However because our knowledge can increase or decrease as a result of the act of measurement, the density matrix or wavefunction which represents this knowledge must change accordingly, subject to the uncertainty principle, but not at all following from a Schr\"odinger equation which might include the apparatus and an observer. Peierls disagrees with Bell that the measurement issues he mentioned are difficult ones.  One gets the impression that these two men may have been arguing since Bell was a student at Birmingham with Peierls in 1956 and they have never come to any resolution.   Peierls also challenges the Landau-Lifshitz  and thus Bohr's view that the apparatus which performs the act of measurement must necessarily be classical i.e. obey classical physics. He cites the example that an experimentalist with good vision can detect down to one photon which can thus constitute a measurement by a non-classical system.  What is classical according to him is that measurements generate a chain of correlated events.  Most of these beyond the first are essentially classical i.e. contain diagonal or near diagonal density matrices.  He was not bothered by Bell's shifty split.  His reasoning in his article\cite{Peierls1991} is a summary of his analysis in his Surprises in Theoretical Physics\cite{Peierls1979} section 1.6.  To him the wavefunction or density matrix represent our knowledge of the system as in the Stern-Gerlach apparatus, since at any stage of a measurement it is in principle possible even after disturbing a system to unscramble it and recover all information until the final change when the observer had registered the result of the measurement.  In this case there is new knowledge and therefore an update of the density matrix becomes necessary.  The concept of any physical system state collapse has no meaning for him.  In fact he was in the process in 1991 of finishing his More Surprises in Theoretical Physics\cite{PeierlsMore1991} when I visited him from Australia. On that occasion he complaint he could not find time to finish it due to many distractions, and the publishers were getting impatient for the manuscript.  In section 1.2 of More Surprises\cite{PeierlsMore1991} he had elaborated a bit more than what was in his 1991 article\cite{Peierls1991} and it was during our conversations about it in reference to what `reality' is that my quote from him above got stuck in my memory. I direct the reader to the opening in section 1.2, ``I have discussed the interpretation of quantum mechanics and emphasized that there is {\it no} alternative to the standard view that the wavefunction (or more generally the density matrix) represents {\it our} knowledge of a physical system" (my italics). One point Peierls did not address was Bell’s “and/or” arguments in his article, which was Bell's strong point ( I agree with Ian Aitchison who raised this to me in private correspondence).  However upon closer examination, this dissection of the shifty split using Boolean logic does not worry me too much now.  I used to worry about it because of my knowledge of the Temple paradox \cite{Temple1935} to which Peierls’ wrote a rebuttal in Nature soon after \cite{Peierls1935}, not entirely satisfactory in my opinion.  However it is interesting that I had noticed Peierls had quite a few books by Betrand Russell on his shelf in his flat during my visits.  They must have both shared pacifism, atheism and perhaps even politics.  Recently I came across the quote generally attributed to Russell:  If you want to study Logic, don’t read Aristotle \cite{BetrandRussell1968}.  Then it clicked.  One just cannot justify applying Boolean logic to a superposition principle of complex valued wavefunctions or density matrices.  So Bell’s “and/or” arguments is no longer such a strong point after all in my humble opinion.

It is worth noting that quantum mechanics just does not deal with individual events, see for example Ballentine \cite{Ballentine1970}. In any individual experiment the outcome is that one will observe a cat to be either alive or dead.  The proposition that Schr\"odinger's superposition describes some kind of pseudo reality with the cat being half dead or half alive with complex phases in Peierls's view is incorrect (see again my above quote).  After all we all know that the Heitler-London model of molecular bonding from the start involves a superposition of this very kind.  The very structure of a lot of chemistry is explained this way. To really question what  is really happening at a deeper level is of course interesting, as we now have eta second pulse lasers and can do experiments that may provide a better picture if indeed quantum mechanics has a deeper structure than we have so far been able to find.  I think this is what Kurt Gottfried \cite{Gottfried1991} meant by telling Bell that a future theory may well involve abstractions and mathematics that make him yearn for good old fashion quantum theory.


\begin{thebibliography}{10}
\bibitem{Peierls1991} Rudolph Peierls, ``In defence of ``measurement"", Physics World, 19-20, January (1991).
\bibitem{FrankClose1} Frank Close, ``TRINITY - The Treachery and Pursuit of the Most Dangerous Spy in History", Allen Lane, Penguin Random Hoise UK (2019).
\bibitem{Bell1990} John Bell, ``Against measurement", Physics World, 33-40, August (1990)
\bibitem{Peierls-Landau1930} L. Landau and R. Peierls, ``Extension of the Uncertainty Principle to Relativistic Quantum Theory", Zeitschrift f\"ur Physik {\bf 69}, 56-69 (1930)
\bibitem{Bohr-Rosenfeld 1950} N. Bohr and L. Rosenfeld, ``Field and Charge Measurements in Quantum Electrodynamics", Physical Review {\bf 78}(6), 794-98, (1950). As early as 1933, Bohr and Rosenfeld and others including Heisenberg in 1934 were seriously trying to resolve the issues raised by Peierls and Landau, see references cited in this paper.
\bibitem{Heisenberg1938} Werner Heisenberg first published his S matrix theory in 1943 in ``Die beobachtbaren Gr\"o\ss  in der Theorie der Elementarteilchen", Zeitschrift f\"ur Physik {\bf 120} 7-10 (1943), though it must have been known as early as 1938, see Landau and Lifshitz \cite{Landau-Lifshitz4}.
\bibitem{Landau-Lifshitz4}   V.B. Berestetskii and E.M. Lifshitz,  Quantum Electrodynamics, Course of Theoretical Physics Vol 4, Pergamon Press, Oxford, UK.
\bibitem{Peierls-Landau21930} L. Landau and R. Peierls, ``Quantum Electrodynamics in Configuration Space", Zeitschrift f\"ur Physik {\bf 62}, 188-200 (1930).
\bibitem{Dirac1963} P. Dirac, ``The Evolution of the Physicist's Picture of Nature", Scientific American {\bf 208}(5), 45-53 (1963).
\bibitem{Peierls1979} R. Peierls, ``Surprises in Theoretical Physics", Princeton University Press, New Jersey (1979).
\bibitem{PeierlsMore1991} R. Peierls, ``More Surprises in Theoretical Physics", Princeton University Press, New Jersey (1991).
\bibitem{Temple1935} G. Temple, ``The Fundamental Paradox of the Quantum Theory", Nature June 8, 957 (1935).
\bibitem{Peierls1935} R. Peierls, ``The Fundamental Paradox of the Quantum Theory", Nature Sep 7, 395 (1935).
\bibitem{BetrandRussell1968} Betrand Russell ``The Art of Drawing Inferences" in The Art of Philosophizing and other essays, New York: Philosophical Library (1968).  See also his ``A History of Western Philosophy", Routledge London 1946.
\bibitem{Ballentine1970} L.E. Ballentine, ``The Statistical Interpretation of Quantum Mechanics ", Rev. Mod. Physics {\bf 42},(4) 358-381 (1970).
\bibitem{Gottfried1991} Kurt Gottfried, ``Does quantum mechanics carry the seeds of its own destruction", Physics World, 34-40, August (1991).

\end{thebibliography}
\end{document}